\documentclass[apjl]{emulateapj}
\usepackage{epsfig}
\usepackage{amsmath,amssymb,bm}
\usepackage[varg]{txfonts}

\begin{document}
\label{firstpage}

\title[Evolution of the CO Snow Line]{On the evolution of the CO snow
  line in protoplanetary disks}

\author{Rebecca G. Martin\altaffilmark{1,3}} \author{Mario
  Livio\altaffilmark{2}} \affil{\altaffilmark{1}JILA, University of
  Colorado \& NIST, UCB 440, Boulder, CO 80309, USA}
\affil{\altaffilmark{2}Space Telescope Science Institute, 3700 San
  Martin Drive, Baltimore, MD 21218, USA }
\affil{\altaffilmark{3}Sagan Fellow}

\begin{abstract}
CO is thought to be a vital building block for prebiotic molecules
that are necessary for life. Thus, understanding where CO existed in a
solid phase within the solar nebula is important for understanding the
origin of life. We model the evolution of the CO snow line in a
protoplanetary disk. We find that the current observed location of the
CO snow line in our solar system, and in the solar system analogue TW
Hydra, cannot be explained by a fully turbulent disk model. With
time-dependent disk models we find that the inclusion of a dead zone
(a region of low turbulence) can resolve this problem. Furthermore, we
obtain a fully analytic solution for the CO snow line radius for late
disk evolutionary times. This will be useful for future
observational attempts to characterize the demographics and predict
the composition and habitability of exoplanets.
\end{abstract}


\section{Introduction}

The most abundant volatiles in a protoplanetary disk are $\rm CO$,
$\rm CO_2$ and $\rm H_2O$. A snow line marks a radial location in a
disk where the mid-plane temperature\footnote{We assume that the gas
  and dust temperatures in the disk are equal. This assumption may
  break down in the upper layers of the disk, but is reasonable at the
  disk midplane \citep[e.g.][]{Dullemond2007}.} drops sufficiently so
that a volatile condenses out of the gas phase to become
solid\footnote{Volatiles condense on to dust grains to become solid
  \citep{Tielens1982} rather than forming pure ice.}. Thus, each
volatile has a different snow line radius, water ice being closest to
the host star, then $\rm CO_2$ and then $\rm CO$. Snow lines are
thought to regulate the planet formation process
\citep[e.g.][]{Oberg2011}. Giant planets, for instance, are
expected to form outside the water snow line because the density of
solids is significantly higher outside of this radius
\citep[e.g.][]{Pollack1996,Morales2011,Ros2013}. Snow lines are also
important because dust piles up in the pressure trap just inside the
snow line and the grains become stickier as ice condenses on their
surfaces. The composition of a planet and its atmosphere are
determined by where the planet forms, and where it accretes material
relative to the snow lines \citep[e.g.][]{Oberg2011}.

The water snow line occurs at a temperature of around $170\,\rm K$
\citep{Hayashi1981} and currently in our solar system is observed to
be at a radius of $2.7\,\rm AU$, within the asteroid belt
\citep{Abeetal2000,Morbidellietal2000,MartinandLivio2013asteroids}. The
CO snow line occurs at a cooler temperature of about $T_{\rm
  CO,snow}=17\,\rm K$ \citep{Oberg2005}.  Comets, from the Kuiper
belt, have varying amounts of CO, suggesting that they formed close to
the CO snow line \citep{Ahearnetal2012}.  Some dwarf planets such as
pluto and comets in the Kuiper belt contain the even more volatile
$\rm N_2$ gas \citep[e.g.][]{Cochran2000} implying that they may have
formed beyond the CO snow line. The Kuiper belt is currently thought
to have formed in the approximate region $27 -35\,\rm AU$
\citep{Levison2008} and thus, at the time of planetesimal formation
the CO snow line would have been in this region. The CO snow line
could mark the transition from the planet forming region to the
formation of smaller icy bodies and dwarf planets, like Pluto.

The water snow line is hard to observe in exosolar systems because it
is very close to the star. However, the CO snow line is an easier
target because it is farther away. The best studied CO snow line is in
a solar nebula analogue disk around TW Hya. The star has a mass of
$0.8\,\rm M_\odot$ and an age of less than $10\,\rm Myr$
\citep{Hoffetal1998}. \cite{Qietal2013} observed the reactive
ion $\rm N_2H^+$ which is only present where CO is frozen out. They
found that the CO snow line lies at $28 -31\,\rm AU$, similar to that
in our solar system.

Understanding the evolution of the CO snow line is essential to
deciphering the origins of prebiotic molecules that are necessary for
life \citep{Tielens1997}. CO ice is needed to form methanol which is a
building block for more complex organic molecules. Comets are thought
to have bombarded the early Earth, thus delivering these molecules to
Earth and allowing life to emerge. In this work we therefore consider
the evolution of the CO snow line in various models of protoplanetary
disks.

Angular momentum transport in protoplanetary disks is thought to be
driven by turbulence generated by the magneto-rotational instability
(MRI) \citep{BH1991}.  However, it is now widely acknowledged that
protoplanetary disks are not sufficiently ionised for the MRI to
operate throughout. They contain a {\it dead zone}, a region of low
turbulence at the disk mid-plane where the MRI is suppressed
\citep[e.g.][]{Gammie1996}.  In the present Letter, we investigate the
evolution of the CO snow line in disks with and without a dead zone.

\section{Fully MRI Turbulent Disk Model}
\label{turb}

Material in an accretion disk orbits the central mass, $M_\star$, with
Keplerian velocity at radius $R$ with angular velocity
$\Omega=\sqrt{GM_\star/R^3}$ \citep[e.g.][]{LP1974,Pringle1981}. The
viscosity in a fully MRI turbulent disk may be parameterised with
\begin{equation}
\nu=\alpha_{\rm m}\frac{c_{\rm s}^2}{\Omega},
\end{equation}
where the \cite{SS1973} viscosity parameter is $\alpha_{\rm m}$, the
sound speed is $c_{\rm s}=\sqrt{{\cal R} T_{\rm c}/\mu}$, ${\cal R}$
is the gas constant, $\mu$ is the gas mean molecular weight and
$T_{\rm c}$ is the mid-plane temperature. The surface density of a
steady state disk is
\begin{equation}
\Sigma =\frac{\dot M}{3\pi\nu}
\end{equation}
\citep{Pringle1981}, where the infall accretion rate is $\dot M$ and
is constant through all radii.  The surface temperature in the steady
disk is
\begin{equation}
\sigma T_{\rm e}^4= \frac{9}{8}\frac{\dot M}{3\pi}\Omega^2+\sigma T_{\rm irr}^4
\label{temp}
\end{equation}
\citep[e.g.][]{Cannizzo1993,Pringleetal1986}. For an unflared disk,
the irradiation temperature is
\begin{equation}
 T_{\rm irr}=\left(\frac{2}{3\pi}\right)^\frac{1}{4}\left(\frac{R_\star}{R}\right)^\frac{3}{4} T_\star
\end{equation}
\citep{Chiang1997}, where $T_\star$ is the temperature and $R_\star$
is the radius of the star.  The mid-plane temperature of the disk is
found from
\begin{equation}
 T_{\rm c}^4=\tau T_{\rm e}^4,
\label{t}
\end{equation}
where the optical depth is
\begin{equation}
\tau =\frac{3}{8}\kappa \frac{\Sigma}{2}
\label{tau}
\end{equation}
and the opacity is
\begin{equation}
\kappa=a T_{\rm c}^b.
\label{kappa}
\end{equation}
Dust dominates the absorption properties of matter where it is
present. Thus, for the low temperatures close to the CO snow line, we
take $a=0.053$ and $b=0.74$ (see \citealt{Zhuetal2009} and also
\citealt{Bell1994}). However, we note that the exact values do not
affect the disk temperature strongly (see equation~\ref{t}).

\begin{figure}
\includegraphics[width=8.4cm]{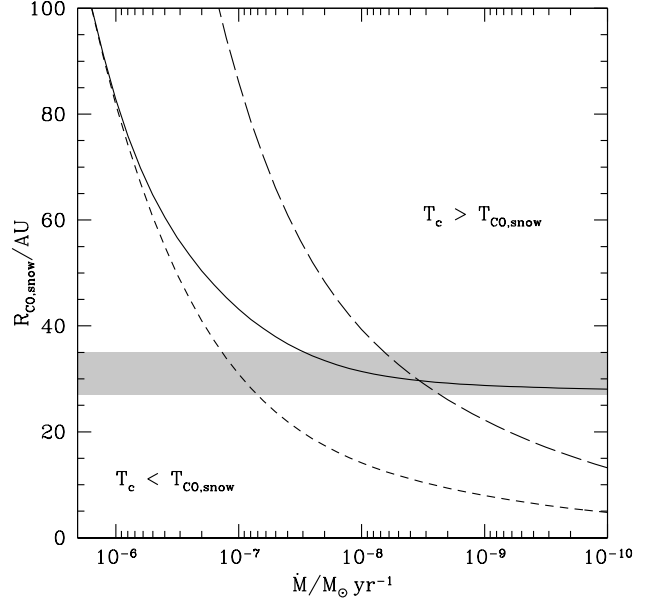}
\caption{The evolution of the CO snow line as a function of accretion
  rate in a steady state disk with $M_\star=1\,\rm M_\odot$, $T_{\rm
    CO,snow}=17\,\rm K$, $T_\star=4000\,\rm K$ and $R_\star=3\,\rm
  R_\odot$. (The accretion rate decreases in time in an evolving
  disk.) The dashed lines show a fully MRI turbulent disk with
  $\alpha_{\rm m}=0.01$ (short-dashed) and $\alpha_{\rm m}=10^{-4}$
  (long-dashed). The solid line shows a disk with a self-gravitating
  dead zone. The shaded region shows the uncertain location of the CO
  snow line in our solar system at the time of planetesimal
  formation. }
\label{coline}
\end{figure}

We solve the equation $T_{\rm c}=T_{\rm CO,snow}$ to find the CO snow
line radius, $R_{\rm CO,snow}$. This radius is shown in
Fig.~\ref{coline} by the short-dashed line as a function of the
accretion rate, $\dot M$, for $M_\star=1\,\rm M_\odot$, $\alpha_{\rm
  m}=0.01$, $T_{\rm CO,snow}=17\,\rm K$, $T_\star=4000\,\rm K$ and
$R_\star=3\,\rm R_\odot$. Because the accretion rate through the disk
drops in time, time is the implicit coordinate here. There is still
some uncertainty concerning the value for $\alpha_{\rm m}$ in
protoplanetary disks \citep[e.g.][]{Kingetal2007} and thus we also
show a disk model with a small $\alpha_{\rm m}=10^{-4}$ by the
long-dashed line.  Given that the CO snow line in our solar system is
thought to have been in the range $27 -35\,\rm AU$ at the time of
planetesimal formation, the CO snow line appears to have the same
evolution problems as the water snow line
\citep{GaraudandLin2007,Okaetal2011,MartinandLivio2012}. That is, in a
fully MRI turbulent disk, the CO snow line moves in too close to the
host star during the low accretion rate phase towards the end of the
disk lifetime. Thus, in the following Section we consider a
time-dependent disk with a dead zone in order to track the evolution
of the CO snow line in a more realistic disk model.

For comparison to water snow line models, we find an analytic fit to
the CO snow line radius. On scales of tens of AU, the irradiation is
certainly the dominant heating source. Thus, we find an approximate
analytical steady state solution by ignoring the viscous heating term
in equation~(\ref{temp}) so that $T_{\rm e}=T_{\rm irr}$.  In this
limit
\begin{align}
R_{\rm CO,snow}   \approx  & \,\,13.2\,  \left(\frac{\alpha_{\rm m}}{0.01}\right)^{-\frac{2}{9}}\left(\frac{M_\star}{\rm M_\odot}\right)^{\frac{1}{9}} \left(\frac{\dot M}{10^{-8}\,\rm M_\odot\, yr^{-1}}\right)^{\frac{2}{9}} \cr & \times \left(\frac{T_{\rm CO,snow}}{17\,\rm K}\right)^{-0.95} 
\left(\frac{R_\star}{3\,\rm R_\odot}\right)^\frac{2}{3}
\left(\frac{T_\star}{4000\,\rm K}\right)^\frac{8}{9}
 \,\rm AU.
\label{rturb}
\end{align}
This is almost identical to the full solution shown in the dashed
lines in Fig.~\ref{coline} for accretion rates $\dot M \lesssim
10^{-8}\,\rm M_\odot\,\rm yr^{-1}$, where irradiation dominates the
viscous heating term. For higher accretion rates, this formula
underestimates the CO snow line radius.

\section{A Disk with a Dead Zone}

When the ionisation fraction is not sufficiently high for the MRI to
drive turbulence, a dead zone forms \citep[see disk structure sketches
  in][]{MartinandLivio2013snow}. The hot inner parts of the disk with
midplane temperature $T_{\rm c}>T_{\rm crit}$ are thermally ionised
and thus MRI active. The value of $T_{\rm crit}$ is thought to be
around $800\,\rm K$ \citep{Umebayashi1988}.  Farther away from the
central star, cosmic rays or X-rays from the star are the dominant
source of ionisation \citep{Glassgoldetal2004} and these can only
penetrate the surface layers with surface density $\le \Sigma_{\rm
  crit}$.  Where the total surface density, $\Sigma$, is larger than
this critical value, $\Sigma>\Sigma_{\rm crit}$, a dead zone exists at
the mid-plane with surface density $\Sigma_{\rm g}=\Sigma-\Sigma_{\rm
  crit}$.  Thus, the MRI is only active in the surface layers. The
precise value of $\Sigma_{\rm crit}$ remains uncertain
\citep[e.g.][]{Martinetal2012a,Martinetal2012b}. If cosmic rays are
the dominant ionisation source, it may be as high as $\Sigma_{\rm
  crit}=200\,\rm g\,cm^{-2}$ \citep{Gammie1996,Fromangetal2002} but if
X-rays dominate the active layer is much smaller
\citep{Matsumura2003}. In the outer parts of the disk, where
$\Sigma<\Sigma_{\rm crit}$, the external ionisation sources penetrate
to the midplane and the disk is fully MRI active. In all parts of the
disk that are MRI active, we assume the same constant viscosity
parameter, $\alpha_{\rm m}$.

Build up of material within the dead zone may cause the disk to become
self gravitating. This occurs when the \cite{Toomre1964} parameter,
$Q=c_{\rm s}\Omega/\pi G\Sigma$, drops below its critical value that
we take to be $Q_{\rm crit}=2$. A second type of turbulence,
gravitational turbulence, is driven with viscosity
\begin{equation}
\nu_{\rm g}=\alpha_{\rm g} \frac{c_{\rm g}^2}{\Omega}.
\end{equation}
We take the \cite{SS1973} parameter to be
\begin{equation}
\alpha_{\rm g}=\alpha_{\rm m} \exp\left(-Q^4\right)
\end{equation}
\citep[e.g.][]{Zhuetal2010b}. However, providing that the function
decreases strongly with $Q$, the form doesn't significantly affect the
viscosity \citep{Zhuetal2010a,Zhuetal2010b,MartinandLubow2013prop}. In
this section we first consider time-dependent numerical models of a
disk with a dead zone and then we find analytic approximations to the
CO snow line radius in such a model.

\subsection{Time-Dependent Protoplanetary Disk Models}

We consider a model for the collapse of a molecular cloud on to the
disk \citep{Armitage2001,Martinetal2012b}. Initially the accretion
rate on to the disk is $2\times 10^{-6}\,\rm M_\odot\, yr^{-1}$ and
this decreases exponentially on a timescale of $10^5\,\rm yr$. The
initial surface density of the disk is that of a fully turbulent
steady disk with an accretion rate of $2\times 10^{-6}\,\rm
M_\odot\,yr^{-1}$ around a star of mass $M_\star=1\,\rm M_\odot$. We
take a radial grid of 200 points evenly distributed in $\log R$ from
$R=1\,\rm AU$ up to $R=200\,\rm AU$ and infalling material is added at
a radius of $R=195\,\rm AU$. The inner boundary has zero torque and
the outer boundary has zero radial velocity.  The CO snow line lies
far from the inner edge of the disk. We take $\alpha_{\rm m}=0.01$ but
note that the chosen value does not significantly affect the CO snow
line evolution for the model with a dead zone. We consider two disk
models, one that is fully MRI turbulent throughout and a second that
has a dead zone determined by $\Sigma_{\rm crit}=10\,\rm
g\,cm^{-2}$. We solve the time-dependent accretion disk equations
along with a simplified energy equation including viscous and
irradiative heating terms \citep[see][for more
  details]{MartinandLubow2011}. We chose $T_\star=4000\,\rm K$,
$R_\star=3\,\rm R_\odot$ and $T_{\rm CO,snow}=17\,\rm K$.

\begin{figure}
\includegraphics[width=8.4cm]{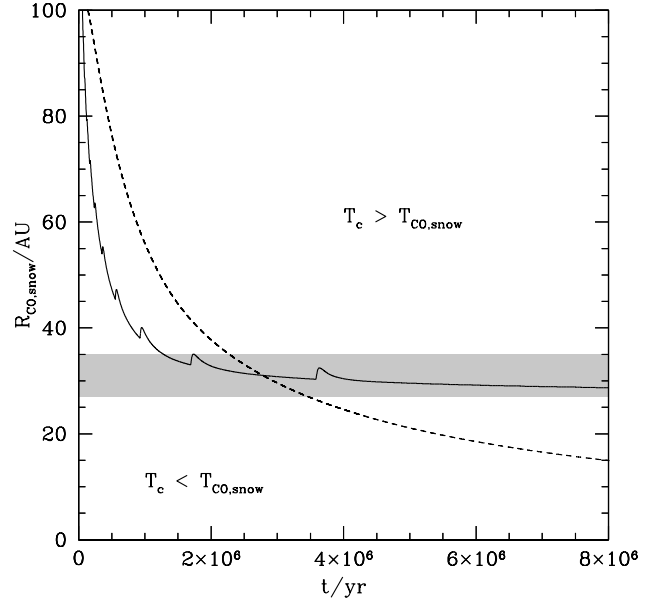}
\caption{Evolution of the CO snow line in a time-dependent disk with
  an exponentially decreasing infall accretion rate. The dashed line
  shows a fully MRI turbulent disk and the solid line a disk with a
  dead zone determined by $\Sigma_{\rm crit}=10\,\rm g\,cm^{-2}$. The
  shaded region shows the location of the CO snow line in our solar
  system at the time of planetesimal formation.}
\label{time}
\end{figure}

The disk is gravo-magneto unstable for large infall accretion
rates. This causes unsteady accretion on to the central star as the
turbulence transitions from gravitationally produced to magnetically
produced \citep[][]{Armitage2001,
  Zhuetal2009,MartinandLubow2011,MartinandLubow2013prop}.  In
Fig.~\ref{time} we show the evolution of the CO snow line as a
function of time for the fully turbulent disk (dashed line) and the
disk with a dead zone (solid line). The model with a dead zone has
small and brief increases in the snow line radius which occur during
the FU Orionis type outbursts. For later times (and smaller accretion
rates), the disk with a dead zone has a CO snow line radius that is
much larger than that of the fully turbulent disk model, and in
agreement with that in our solar system. A dead zone therefore appears
to be a necessary component in modeling protoplanetary disks. With a
dead zone included, we have shown that time-dependent numerical
simulations predict a larger CO snow line radius (in agreement with
the observations) because the small amount of self-gravity within the
dead zone heats the more massive disk.

\subsection{Analytical Solutions}
\label{dz}

We have shown in the previous section that the presence of a dead zone
significantly affects the evolution of the CO snow line. Once the
infall accretion rate drops sufficiently, the outbursts cease but a
dead zone may still be present.  Following
\cite{MartinandLivio2013snow}, because the CO snow line is in the
self-gravitating part of the dead zone, we find steady state analytic
solutions for its radius.  The solution has MRI active surface layers
(with surface density $\Sigma_{\rm crit}$) over a self-gravitating
dead zone. In order to find analytic solutions, we work in the limit
$\Sigma \gg \Sigma_{\rm crit}$ and approximate $\Sigma_{\rm crit}=0$.

When the disk is self gravitating, it has surface density
\begin{equation}
\Sigma = \frac{c_{\rm g}\Omega}{\pi G Q}.
\label{sig}
\end{equation}
For a steady state accretion disk the accretion rate is
\begin{equation}
\dot M  = 3 \pi \nu_{\rm g}\Sigma .
\end{equation}
Both $\Sigma$ and $\nu_{\rm g}$ depend on $Q$, and thus we can relate
the Toomre parameter to the accretion rate through
\begin{equation}
 \dot M = \left(\frac{3 \, c_{\rm g}^3\alpha_{\rm m}}{G}\right) \frac{\exp\left(-Q^4\right) }{Q}.
\label{qq}
\end{equation}
The term in brackets is constant for a fixed CO snow line
temperature. This expression depends sensitively on $Q$ and thus for a
reasonable range of accretion rates, $Q$ is approximately constant
\citep[see also][]{MartinandLivio2012}.  We scale the variables to
$T_{\rm CO,snow}'=T_{\rm snow}/17\,\rm K$, $\alpha_{\rm
  m}'=\alpha_{\rm m}/0.01$, $M_\star'=M_\star/\rm M_\odot$ and $R'=R/\rm AU$ and
solve equation~(\ref{qq}) to find the scaled Toomre parameter
\begin{equation}
Q'=\frac{Q}{Q_{\rm crit}}=0.69\,\left[ \frac{W(x)}{W(x_0)}\right]^\frac{1}{4},
\label{q}
\end{equation}
where we define
\begin{equation}
x=2.45\times 10^{7} \frac{\alpha_{\rm m}'^4 T_{\rm CO,snow}'^6}{\dot M_\star'^4}
\end{equation}
and $x_0=2.45\times 10^{7}$. The Lambert function, $W$, is defined by
the equation
\begin{equation}
x=W(x)\exp[W(x)].
\end{equation}

The mid-plane temperature is related to the disk surface temperature
through equations~(\ref{t})--(\ref{kappa}).  The steady energy
equation is
\begin{equation}
\sigma T_{\rm e}^4=\frac{9}{8}\nu_{\rm g} \Sigma \Omega^2+\sigma T_{\rm irr}^4.
\label{temp2}
\end{equation}
We solve $T_{\rm c}=T_{\rm CO,snow}$ to find the CO snow line radius
and plot it as a function of the accretion rate. This is shown by the
solid line in Fig.~\ref{coline} for $M_\star=1\,\rm M_\odot$, $T_{\rm
  CO,snow}=17\,\rm K$, $\alpha_{\rm m}=0.01$, $T_\star=4000\,\rm K$
and $R_\star=3\,\rm R_\odot$. As shown by the numerical models, for
small accretion rates, the CO snow line radius is insensitive to the
accretion rate through the disk. Thus we can find this radius
analytically.

In the limit where the irradiation is the dominant heating source (for
low accretion rates $\dot M\lesssim 10^{-8}\,\rm M_\odot\,yr^{-1}$) we
approximate equation~(\ref{temp2}) with $T_{\rm e}=T_{\rm irr}$. This
gives the analytic CO snow line radius
\begin{equation}
R_{\rm CO,snow}\,\,  =\,\,  29.3\,{M_\star'}^{\frac{1}{9}}R_\star'^{\frac{2}{3}}T_\star'^{\frac{8}{9}}  T_{\rm CO,snow}'^{-0.61}   \left[\frac{W(x)}{W(x_0)}\right]^{-\frac{1}{18}}     \,\rm AU.
\label{full}
\end{equation}
  Its value is given
approximately by
\begin{align}
R_{\rm CO,snow} \approx &  29.3 \,
\left(\frac{M_\star}{\rm M_\odot}\right)^{\frac{1}{9}}
\left(\frac{R_\star}{3\,\rm R_\odot}\right)^{\frac{2}{3}}
\left(\frac{T_\star}{4000\,\rm K}\right)^{\frac{8}{9}} \cr & \times
\left(\frac{T_{\rm CO,snow}}{17\,\rm K}\right)^{-0.61} \,\rm  AU.
\label{app}
\end{align}
This is almost identical to the solid line shown in Fig.~\ref{coline}
for low accretion rates. Note that this is independent of $\alpha_{\rm m}$.

\begin{figure}
\includegraphics[width=8.4cm]{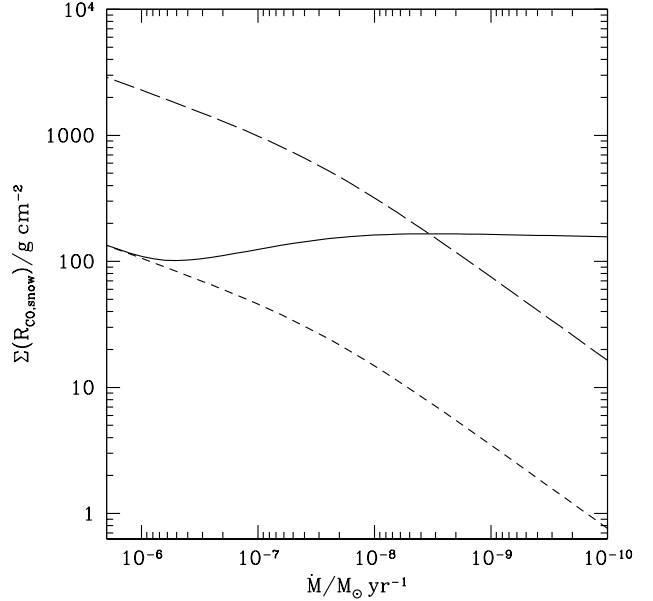}
\caption{The surface density at radius $R=R_{\rm CO,snow}$ (as shown
  in Fig.~\ref{time}) for the steady state CO snow line solutions
  including both viscous and irradiative heating with $M_\star=1\,\rm
  M_\odot$, $\alpha_{\rm m}=0.01$, $T_{\rm CO,snow}=17\,\rm K$,
  $T_\star=4000\,\rm K$ and $R_\star=3\,\rm R_\odot$. The dashed lines
  are the fully turbulent disk model with $\alpha_{\rm m}=0.01$ (short-dashed)
  and $\alpha_{\rm m}=10^{-4}$ (long-dashed). The solid line is the disk with
  a self-gravitating dead zone. }
\label{sigma}
\end{figure}

In Fig.~\ref{sigma} we show the surface density of the steady state
solutions at the CO snow line radius (including both viscous and
irradiative heating terms). The disk with a dead zone has a fairly
constant surface density at the snow line radius of around $100\,\rm
g\,cm^{-2}$. Thus, provided that $\Sigma_{\rm crit} \ll 100\,\rm
g\,cm^{-2}$, then the solutions presented in this section are
valid. As shown in the previous section, $\Sigma_{\rm crit}=10\,\rm
g\,cm^{-2}$ was small enough for the dead zone to persist at the CO
snow line radius for longer than the lifetime of the disk.

Submillimeter observations of TW~Hya combined with radiative transfer
calculations of the disk structure that assume a constant dust to gas
ratio with radius predict a surface density about an order of
magnitude lower than that required in our models
\citep{Andrews2012}. However, their models were unable to reproduce
both the brightness profiles and the CO line emission. It is possible
that with the inclusion of a dead zone in their disk models that the
observed features may be reproduced. The dead zone could explain why
the dust emission has a sharp outer edge at $60\,\rm AU$ while the CO
emission extends out past $215\,\rm AU$ and this should be
investigated in future work.

\section{Discussion and Conclusions}
\label{discussion}

Shearing box simulations suggest that a small amount of turbulence may
be driven in the dead zone by the turbulence in the disk surface
layers \citep[e.g.][]{Fleming2003, Simonetal2011}.  The addition of
this small viscosity is unlikely to suppress outbursts
\citep{MartinandLubow2013dza}, although the triggering may be due to
heating from the additional turbulence, rather than self gravity
\citep{Baeetal2013}.  In the limit of small active layer surface
density, the steady state dead zone solution would be the same as the
fully MRI turbulent solution in Section~\ref{turb} but with a smaller
$\alpha_{\rm m}$. In the figures we have also considered a smaller
turbulence of $\alpha_{\rm m}=10^{-4}$ for this comparison. We find
that, with an $\alpha_{\rm m}$ that is two orders of magnitude smaller
than that in the active layers, it remains difficult to explain the
current location of the CO snow line in our solar system, and in TW
Hya.

The CO snow line in the disk around the Herbig~Ae star HD 163296 has
been found to lie at a radius of around $155\,\rm AU$
\citep{Qietal2011}. With observed parameters of $M_\star=2.3\,\rm
M_\odot$, $T_\star=9333\,\rm K$, $R_\star=2\,\rm R_\odot$ and $\dot
M=7.6\times 10^{-8}\,\rm M_\odot\,yr^{-1}$, our fully turbulent disk
model (equation~\ref{rturb}) predicts a CO snow line radius of
$37\,\rm AU$. The dead zone model (equation~\ref{app}) predicts a
radius of $62\,\rm AU$. Thus, neither model can explain such a large
CO snow line radius (although the model with a dead zone shifts the
radius in the right direction). We suggest that disk flaring could
account for this. Approximations for the temperatures of flared disks
would predict a CO snow line radius $>100\,\rm AU$ \citep[see Fig. 4
  in][]{Chiang1997}. In contrast, the inner regions of the disk in
TW~Hya are flat, although it is moderately flared at radii $R>45\,\rm
AU$. Thus, flaring does not affect the CO snow line radius in
TW~Hya. Furthermore, this would suggest that our solar nebula was not
flared, at least in the inner regions.

%With ALMA, the CO snow line radius will be resolved in many more
%systems in the near future. The work presented here predicts where the
%CO snow line will be found depending on the observed properties of the
%system.  Furthermore, properties and compositions of exoplanets may be
%predicted given the location of the planet relative to the snow
%lines. Being able to predict the location of the snow lines in a
%system may therefore be important for future attempts to characterize
%the demographics of exoplanets.

%Snow lines model the planet forming regions of protoplanetary
%disks. They determine where different types of planets, such as gas
%giants and terrestrial, may form. The CO snow line is particularly
%important for determining the regions of the protoplanetary system
%from which prebiotic molecules originate. In our solar system, these
%molecules, that formed the building blocks of life, were probably
%delivered to Earth after its formation in the form of comets and
%asteroids. Thus, a Kuiper belt equivalent that exists beyond the CO
%snow line may be necessary for a planet to harbour life.

We have found that a fully MRI turbulent disk predicts a CO snow line
that is much closer to the host star than that observed in our solar
system and in the solar neubula analogue TW Hya. With a dead zone, a
small amount of self-gravity heats the more massive disk and the CO
snow line radius is moved outwards (in agreement with the
observations). We have also found a fully analytic solution for the
snow line radius in a disk with a dead zone for low infall accretion
rates, appropriate for the later stages of protoplanetary disk
evolution. The solution is valid providing that the surface density
ionised by external sources, $\Sigma_{\rm crit} \ll 100\,\rm
g\,cm^{-2}$. This formula could prove useful for determining
composition and habitability of exo-solar planets.

\section*{Acknowledgments} 
We thank an anonymous referee for comments that have improved the
manuscript. RGM's support was provided under contract with the
California Institute of Technology (Caltech) funded by NASA through
the Sagan Fellowship Program.

\bibliographystyle{apj}

\label{lastpage}
\end{document}